\documentstyle[12pt]{article}
\input epsf

\begin{document}

\begin{titlepage}

\begin{flushright}
Freiburg--THEP 96/14\\
July 1996
\end{flushright}
\vspace{2.5cm}

\begin{center}
\large\bf
{\LARGE\bf Two--loop effects of enhanced electroweak strength 
           in the Higgs sector \footnote{Contributed to the 
           {\em Quarks '96} conference, 
           5---11 May 1996, Yaroslavl, Russia}}\\[2cm]
\rm
{Adrian Ghinculov}\\[.5cm]

{\em Albert--Ludwigs--Universit\"{a}t Freiburg,
           Fakult\"{a}t f\"{u}r Physik}\\
      {\em Hermann--Herder Str.3, D-79104 Freiburg, Germany}\\[3.5cm]
      
\end{center}
\normalsize

\begin{abstract}

The selfcoupling of the Higgs field grows with the mass of the Higgs
particle and induces potentially large radiative corrections in the
electroweak model. The technical aspects of performing multiloop calculations
in the massive case are discussed briefly. I review the status of two--loop 
calculations of radiative corrections of enhanced electroweak strength 
which are relevant for the Higgs physics. I discuss the relevance of the 
existing results with respect to heavy Higgs searches at future colliders 
and their implications regarding the validity range of perturbation theory.
\end{abstract}

\vspace{3cm}

\end{titlepage}


\title{Two--loop effects of enhanced electroweak strength in the Higgs sector}

\author{Adrian Ghinculov}

\date{{\em Albert--Ludwigs--Universit\"{a}t Freiburg,
           Fakult\"{a}t f\"{u}r Physik},\\
      {\em Hermann--Herder Str.3, D-79104 Freiburg, Germany}}

\maketitle

\begin{abstract}

The selfcoupling of the Higgs field grows with the mass of the Higgs
particle and induces potentially large radiative corrections in the
electroweak model. The technical aspects of performing multiloop calculations
in the massive case are discussed briefly. I review the status of two--loop 
calculations of radiative corrections of enhanced electroweak strength 
which are relevant for the Higgs physics. I discuss the relevance of the 
existing results with respect to heavy Higgs searches at future colliders 
and their implications regarding the validity range of perturbation theory.
\end{abstract}


\section{Introduction}

The Higgs resonance required by the simplest version 
of a spontaneous electroweak symmetry breaking sector 
has eluded so far a direct detection. Indirect measurements 
of the Higgs mass still give rather loose bounds because 
of screening of the Higgs effects in low energy radiative 
corrections. In fact, a minimal Higgs with a mass of the order 
of 1 TeV is not excluded by the available data \cite{weiglein}. Still, the 
possibility of a heavy Higgs boson raises both theoretical 
and phenomenological problems because the selfinteraction of 
the Higgs field increases with the Higgs mass.

How does the Higgs sector behave in the strong coupling regime? 
This is an interesting question for which no definite answer 
exists yet. A number of approaches were proposed, although each 
has its own problems. An idea suggested long time ago by Veltman 
implies the formation of bound states of weak bosons which would 
behave like Higgs bosons with enhanced couplings to the vector 
bosons and to themselves \cite{veltman}. 
For gaining insight into the behaviour 
of the Higgs sector at large selfcouplings, the nonperturbative 
$1/N$ expansion technique was developed. These models can be solved 
easily at leading order, and reveal for instance an interesting 
relation between the Higgs mass and width which deviates from the 
perturbative result for large couplings \cite{einhorn}. In particular, 
the Higgs mass saturates at a value of the order of 900 GeV 
when the quartic coupling is increased. Unfortunately, calculations 
beyond the leading order in the $1/N$ expansion are technically 
extremely difficult \cite{root}. Also these models suffer of pathologies 
associated with the possible triviality of the $\phi^4$ theory 
when treated as renormalized fundamental theories. These problems 
show up in the presence of tachyons in the spectrum of the theory 
at leading order, and lead to problems in the higher orders of the 
$1/N$ expansion. To avoid this kind of inconsistencies, 
one can treat the theory 
as an effective theory by including explicitly a cutoff \cite{schnitzer}.

Setting upper bounds on the Higgs mass by requiring that the 
Higgs boson mass be lower than the triviality scale was attempted 
by lattice calculations \cite{hasenfratz}. 
These results must be interpreted with 
caution because the bounds obtained are regularization 
dependent -- for instance they vary with the geometry of 
the lattice \cite{heller}. 
On the other hand, the top quark may play quantitatively a 
r\^ole in the triviality issue because of its large Yukawa coupling, 
as suggested by the position of the Landau pole. Unfortunately,
it is not straightforward to include these effects in a lattice 
calculation because of well--known difficulties with treating 
fermions on the lattice \cite{stephenson}. 
Moreover, the triviality of the $\phi^4$ theory is still an open question,
and it is not clear that the gauged version of the sigma model is trivial
after all. For a discussion of the triviality of the $\phi^4$ theory
see for instance ref. \cite{callaway}.
This sheds doubts on the relevance of the 
triviality bounds on the Higgs mass 
set by lattice Monte Carlo simulations of the $\phi^4$ theory.

Before nonperturbative effects enter the scene, there is 
a regime where perturbation theory still can be used in the 
Higgs sector, but the problems related to the divergence of 
the perturbation series start to show up in the form of large 
radiative corrections, unitarity violations and large 
renormalization scheme dependency. These effects are transmitted 
to the gauge sector because of the equivalence theorem.
For even larger couplings, the perturbation theory totally breaks 
down, and one cannot rely anymore on perturbative results.
This raises a number of questions in view of the heavy Higgs searches 
at future colliders. At present all phenomenological studies of the Higgs
production and decay mechanisms at future colliders are based on perturbation
theory. One would like to know which is the reliability range of these
predictions, how large are the theoretical uncertainties of the calculation,
and how far in the loop expansion it makes sense to go for improving 
the result. 

To study such higher order effects, one needs techniques to deal 
with massive multiloop Feynman diagrams. Considerable progress has 
been made recently in handling multiloop diagrams at two--, three-- and 
four--loop order in QCD \cite{qcd}. Still, the massive case is technically 
much more difficult, and calculations of physical processes at 
two--loop level were not available until recently, in spite of 
the considerable effort devoted to solving massive two--loop diagrams. 
The reason for this is that such diagrams are in general very 
complicated functions which cannot be expressed analytically in 
terms of usual functions -- see for instance ref. \cite{lauricella}. 
Therefore one has to rely at least partly 
on special techniques for performing calculations of physical relevance.

With the advent of powerful methods for treating such diagrams, 
a number of physical processes involving the scalar sector were 
calculated recently at two--loop level, and provide more insight 
into  the structure of electroweak radiative corrections at large 
Higgs selfcoupling. I review in this talk the status of these 
calculations and their significance for heavy Higgs searches at the LHC.


\section{The model and the techniques}

The leading $m_H$ electroweak corrections to processes which
involve the symmetry breaking scalars at energies not negligible 
when compared to the Higgs mass can grow as $m_H^2$ in the one--loop
approximation and as $m_H^4$ at two--loop. Of course, the leading 
contributions must cancel in the low energy limit due to 
the screening theorem. 

The evaluation of the leading $m_H$ electroweak corrections can be greatly 
simplified by using the equivalence theorem and by working in Landau gauge,
as it was noticed in \cite{marciano}, where this 
scheme was used for calculating the Higgs decay width into longitudinal 
vector bosons at one--loop.
By counting powers of $m_H$, it follows then that the only contributions 
of the desired order come from the diagrams which contain only scalars. 
Therefore it suffices to consider only the sigma model Lagrangian of 
the Higgs sector:

\begin{eqnarray}
{\cal L} & = & 
\frac{1}{2} (\partial_{\mu}H_{0})(\partial^{\mu}H_{0}) +
\frac{1}{2} (\partial_{\mu}z_{0})(\partial^{\mu}z_{0}) +
            (\partial_{\mu}w_{0}^{+})(\partial^{\mu}w_{0}^{-}) 
                                                \nonumber \\
& & - g^{2}\frac{m_{H_{0}}^{2}}{m_{W_{0}}^{2}} \frac{1}{8} \,
[ \, w_{0}^{+} w_{0}^{-} + \frac{1}{2} z_{0}^{2} + \frac{1}{2} H_{0}^{2}
+ \frac{2 m_{W_{0}}}{g} H_{0} 
+ \frac{4 \, \delta t}{g^{2} \, \frac{m_{H_{0}}^{2}}{m_{W_{0}}^{2}}}
 \, ]^{2}
      \; \; ,
\end{eqnarray}
where 
$m_{H_{0}}^{2} = m_{H}^{2} - \delta m_{H}^{2}$,
$m_{W_{0}}^{2} = m_{W}^{2} - \delta m_{W}^{2}$
are bare masses, and 
$H_{0} = Z_{H}^{1/2} H$,
$z_{0} = Z_{G}^{1/2} z$,
$w_{0} = Z_{G}^{1/2} w$
are bare fields. The tadpole counterterm $\delta t$ is determined 
by the condition that the 
Goldstone bosons remain massless and that the vacuum expectation value
of the Higgs field $v$ does not receive quantum corrections.
One can define the gauge coupling constant $g$ at low energy
by using the muon decay as $g^{2} = 4 \sqrt{2} \, m_{W}^{2} \, G_{F}$, 
with $G_{F} = 1.16637 \cdot 10^{-5} \; GeV^{-2}$, and $m_{W} = 80.22 \; GeV$.
The gauge coupling constant is not renormalized at leading order in $m_H$.  

Most phenomenological studies related to heavy Higgs searches at future
colliders use the OMS renormalization scheme, and this is the scheme 
adopted in the following as well. One imposes the following renormalization
conditions to fix the counterterms at the desired order:

\begin{eqnarray}
 & & \hat{\Sigma}_{HH} (k^2=m_H^2) + i \, \delta m_H^2 - i \, \delta t 
+ i \, \delta m_H^2 \, \delta Z_H - i \, \delta t \, \delta Z_H   
 =  0  \nonumber \\
 & & \frac{\partial}{\partial k^2} \hat{\Sigma}_{HH} (k^2=m_H^2) 
 + i \, \delta Z_H  =  0      \nonumber \\
 & & \hat{\Sigma}_{w^+ w^-} (k^2=0) 
- i \, \delta t - i \, \delta t \, \delta Z_G  =  0      \nonumber \\
 & & \frac{\partial}{\partial k^2} \hat{\Sigma}_{w^+ w^-} (k^2=0) 
+ i \, \delta Z_G   =  0     \nonumber \\
 & & \hat{\Sigma}_{W^+ W^-} (k^2=0) + i \, \delta m_W^2   =  0   
      \; \; .
\end{eqnarray}
In this notation, the selfenergies $\hat{\Sigma}$ contain the 
loop and loop--counterterm selfenergy diagrams, but not the 
pure counterterm diagrams. 

It is straightforward to evaluate eqns. 2 at one--loop order
and to determine the one--loop counterterms. One has to keep
in mind that for carrying out a calculation at two--loop order
by using the dimensional regularization, the one--loop 
counterterms are needed up to order $\epsilon$. These
terms result in finite contributions at two--loop order when 
combined with $1/\epsilon$ poles.

In order to determine the counterterms at two--loop order, one
has to calculate two--loop selfenergy diagrams of the topologies shown
in fig. 1. It is well--known that all two--loop diagrams with zero 
external momenta are expressible analytically in terms of Spence functions
\cite{vdbij:rho}.
This is the case with the Goldstone and the vector boson selfenergies.
If one needs to evaluate the diagrams at finite external momenta -- as
is the case with the Higgs selfenergy -- this is in general 
not possible anymore. For these cases a method was developed in ref. 
\cite{2loop:method} which is a hybrid of analytical and numerical techniques.
It can be used to treat any two--loop diagram with arbitrary internal
masses and for arbitrary external momenta.

\begin{figure}
\hspace{1.5cm}
    \epsfxsize = 12cm
    \epsffile{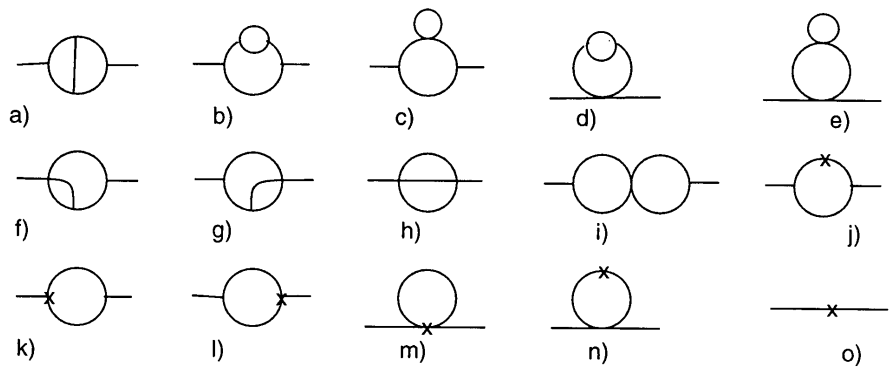}
\caption{The topologies of the two--loop selfenergy diagrams.}
\end{figure}

In fact, the problem at hand is considerably 
simpler because it is essentially a one--scale problem. The Goldstone
mass is zero in Landau gauge, and the only mass scale is the Higgs mass.
The nonvanishing external momentum is also $s=m_H^2$ because of the
on--shell renormalization scheme. This is a considerable simplification when
compared with the general mass case, and indeed in ref. \cite{maher} it was
possible to evaluate most diagrams involved in the calculation
of the on--shell scalar selfenergies
analytically. Also working in lower dimensions can make the
calculation simpler. For instance, in ref. \cite{rajantie} 
the Higgs selfenergy was derived 
completely analytically at too--loop order as a function of the external
momentum in three dimensions. 

However, in more complicated cases, like
for instance more external legs or arbitrary internal masses, numerical
approaches seem unavoidable in realistic calculations. Details of 
such a general approach were described in 
ref. \cite{2loop:method,2loop:Htoww}. Here I will 
only give a sketch of these techniques. The main idea is that any
two--loop diagram can be expressed in terms of two basic scalar integrals
${\cal F}$ and ${\cal G}$, which are defined as follows: 

\begin{eqnarray}
\lefteqn{{\cal G}(m_{1},m_{2},m_{3};k^2) \, \equiv}  \nonumber \\
& & \int d^{n}p\,d^{n}q\, 
       \frac{1}{
             (p^{2}+m_{1}^{2})^{2} \,
             [(q+k)^{2}+m_{2}^{2}] \,
             [(p+q)^{2}+m_{3}^{2}]
	    } \,  =  \nonumber \\
& &   \pi^{4} \{ \,   \frac{2}{\epsilon^{2}} 
    + \frac{1}{\epsilon} [- 1 + 2 \gamma + 2 \log (\pi \, m_{1}^{2}) ]
    + \frac{1}{4} + \frac{\pi^{2}}{12} 
                                                \nonumber \\  
& & + \frac{1}{4} [- 1 + 2 \gamma + 2 \log (\pi \, m_{1}^{2}) ]^{2}
    - 1 + g(m_{1},m_{2},m_{3};k^2)
          \,    \}  + {\cal O}(\epsilon) \; \; ,     \\
\lefteqn{{\cal F}(m_{1}, m_{2}, m_{3} ;k^2) \, \equiv}  \nonumber \\
& & - \int d^{n}p\,d^{n}q\, 
       \frac{(p+q).k}{
             (p^{2}+m_{1}^{2})
             [(q+k)^{2}+m_{2}^{2}]
             (r^{2}+m_{3}^{2})^{2}
	    } \, =  \nonumber \\
& & k^{2} \pi^{4} \{ \, - \frac{1}{2 \epsilon} 
                     + \frac{9}{8}  
    - \frac{1}{2} [ \gamma + \log (\pi m_{1}^{2}) ]
    + f(m_{1},m_{2},m_{3};k^2) 
          \,    \}  + {\cal O}(\epsilon) \; \; .
\end{eqnarray}

The finite parts $f$ and $g$ of the scalar integrals ${\cal F}$ and ${\cal G}$
cannot be expressed in the general mass case in terms of usual functions,
although it may be possible to relate them to the Lauricella function. For
evaluating these functions numerically with high accuracy in an efficient way
it is more convenient to use the following one--dimensional integral
representations:

\begin{eqnarray}
g(m_{1},m_{2},m_{3};k^2) & = &   \int_{0}^{1}\,dx\,  
     [ \, Sp(\frac{1}{1-y_{1}}) + Sp(\frac{1}{1-y_{2}}) 
                                                \nonumber \\  
& &  + y_{1}\log \frac{y_{1}}{y_{1}-1} + 
      y_{2}\log \frac{y_{2}}{y_{2}-1} \, ]
      \; \; ,    \\ 
f(m_{1},m_{2},m_{3};k^2) & = &  \int_{0}^{1}\,dx\,      
     [ \, \frac{1-\mu^{2}}{2 \kappa^{2}}   
                                                \nonumber \\
& & - \frac{1}{2} \, y_{1}^{2} \, \log \frac{y_{1}}{y_{1}-1}  
    - \frac{1}{2} \, y_{2}^{2} \, \log \frac{y_{2}}{y_{2}-1} \,  ]
      \; \; ,
\end{eqnarray}
where the following notations were introduced:

\begin{eqnarray}
y_{1,2} & = & \frac{1 + \kappa^{2} - \mu^{2}
                    \pm \sqrt{\Delta}}{2 \kappa^{2}}  \nonumber \\
\Delta  & = & (1 + \kappa^{2} - \mu^{2})^{2} 
          + 4 \kappa^{2} \mu^{2} - 4 i \kappa^{2} \eta 
      \; \; ,   \nonumber \\
%
   \mu^{2}  & = &  \frac{a x + b (1-x)}{x (1-x)}   \nonumber \\
         a  & = &  \frac{m_{2}^{2}}{m_{1}^{2}} \, , \; \; \; \;
         b \; = \; \frac{m_{3}^{2}}{m_{1}^{2}} \, , \; \; \; \;
\kappa^{2} \; = \; \frac{    k^{2}}{m_{1}^{2}} 
      \; \; .
\end{eqnarray}

After continuing the integrands at complex values of the Feynman
parameter $x$ and carefully inspecting the analytical properties of these
functions, one is able to define an optimized integration path in
terms of spline functions along which the numerical integration 
can be performed very efficiently. The evaluation of these functions
by numerical integration with an accuracy of 8 digits 
takes typically about 50 ms on an HP Apollo 9000/720 workstation.

This technique was
used in ref. \cite{2loop:method,2loop:Htott} 
to calculate the counterterms of eqns. 1 at
two--loop order. A complete list of the one-- and two--loop counterterms 
can be found for instance in ref. \cite{2loop:Htoww}. 
They agree with the results of ref. \cite{maher}, which used different methods
and a slightly different definition of the counterterms. Recently a similar
technique was proposed for calculating a certain class of massive three--loop
Feynman diagrams efficiently \cite{3loop}, but unfortunately 
at present there is no general solution
which would allow one to deal in a systematic way with all possible topologies
of massive three--loop diagrams.


\section{Heavy Higgs decays at two--loop order}

Heavy Higgs bosons mainly decay into pairs of longitudinal vector bosons
and into $t \bar{t}$ pairs. At leading order, these decay widths are given by
the following expressions:

\begin{eqnarray}
\Gamma^{(tree)}_{H \rightarrow t \bar{t}} & = & 
 \frac{3 g^2}{32 \pi} \frac{m_H \, m_t^2}{m_W^2} 
 \left[ 1 - 4 \frac{m_t^2}{m_H^2} \right]^{3/2}  \; \; ,
    \nonumber \\
\Gamma^{(tree)}_{H \rightarrow W^+ W^-} & = & 
 \frac{g^2}{64 \pi} \frac{m_H^3}{m_W^2} 
 \left[ 1 - 4 \frac{m_W^2}{m_H^2} \right]^{1/2}  
 \left[ 1 - 4 \frac{m_W^2}{m_H^2} + 12 \frac{m_W^4}{m_H^4} \right] \; \; ,
    \nonumber \\
\Gamma^{(tree)}_{H \rightarrow Z^0 Z^0} & = & 
 \frac{g^2}{128 \pi} \frac{m_H^3}{m_W^2} 
 \left[ 1 - 4 \frac{m_Z^2}{m_H^2} \right]^{1/2}  
 \left[ 1 - 4 \frac{m_Z^2}{m_H^2} + 12 \frac{m_Z^4}{m_H^4} \right]
      \; \; .
\end{eqnarray}

For large $m_H$, the decays are affected by potentially large electroweak
corrections of order $\lambda = (\frac{g}{4 \pi} \frac{m_{H}}{m_{W}})^2$ 
at one--loop order, and of order $\lambda^2$ at two--loop.

The leading $m_H$ radiative corrections to the $H \rightarrow t\bar{t}$
decay come entirely from the counterterm contributions, and are given
by a correction factor 
$Z_{H} / (1-\frac{\delta m_{W}^{2}}{m_{W}^{2}})$.
Triangle vertex diagrams do not contribute at the order considered here because
these diagrams contain additional powers of the quark mass.
Therefore the leading $m_H$ corrections to the top decay width of the Higgs
boson require only the evaluation of selfenergy diagrams. These corrections
were calculated at two--loop order in ref. \cite{2loop:Htott} with the methods described
in the previous section, and agree with ref. \cite{kniehl} which uses different 
methods:

\begin{eqnarray}
\lefteqn{ \Gamma_{H \rightarrow t \bar{t}} \, =  \,
                      \Gamma^{(tree)}_{H \rightarrow t \bar{t}} \, \times 
     \left[
 1 + \lambda 
     \left(  \frac{13}{8} - \frac{\pi \, \sqrt{3}}{4} \right)  
   - \lambda^2
     \left( \, .51023 \pm 2.5 \cdot 10^{-4} \, \right)
 \right]
 }
 \nonumber \\
 & = & \Gamma^{(tree)}_{H \rightarrow t \bar{t}} \,
  \left[
 1 + .264650 \, \lambda 
   - \left( \, .51023 \pm 2.5 \cdot 10^{-4} \, \right) \lambda^2
 \right]
      \; \; .
\end{eqnarray}

For calculating the leading $m_H$ radiative corrections to the Higgs decay
into vector bosons in a simple way, one can use the equivalence theorem 
and replace the external vector bosons by the corresponding Goldstone bosons.
The one--loop result was derived for instance in ref. \cite{marciano}. 
For extending this
result at two--loop order, one has to calculate the diagrams shown in fig. 2.
This was done in ref. \cite{2loop:Htoww} 
by using the methods described in the previous 
section, and the result reads:

\begin{figure}
\hspace{2.5cm}
    \epsfxsize = 10cm
    \epsffile{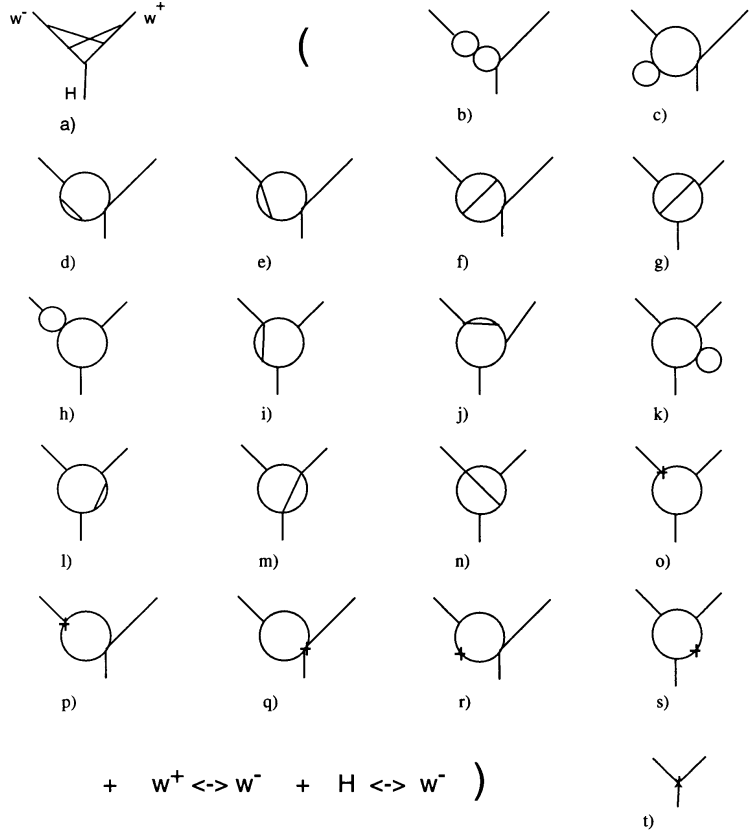}
\caption{The topologies of the vertex diagrams which contribute to 
         the $H \rightarrow ww$ decay at two--loop order.}
\end{figure}

\begin{eqnarray}
\lefteqn{ \Gamma_{H \rightarrow W^+ W^- \, , \, Z^0 Z^0} \, =  \,
          \Gamma^{(tree)}_{H \rightarrow W^+ W^- \, , \, Z^0 Z^0} \, \times }
 \nonumber \\
 & &  \left[
 1 + \lambda 
     \left( \frac{19}{8} + \frac{5 \, \pi^2}{24} 
          - \frac{3 \, \sqrt{3} \, \pi}{4} 
    \right)  
   + \lambda^2
     \left( \, .97103 \pm 8.2 \cdot 10^{-4} \, \right)
 \right]
    \nonumber \\
 & = & \Gamma^{(tree)}_{H \rightarrow W^+ W^- \, , \, Z^0 Z^0} \,
 \left[
 1 + .350119 \, \lambda 
   + \left( \, .97103 \pm 8.2 \cdot 10^{-4} \, \right) \lambda^2
 \right]
      \; \; .
\end{eqnarray}

This result was confirmed very recently by an independent calculation by
A. Frink, B.A. Kniehl, D. Kreimer and K. Riesselmann   \cite{frink},
who used different methods for the evaluation of the two--loop integrals
which are involved.

Of course, in eqns. 9 and 10 some incomplete 
subleading contributions are present in 
the radiative corrections. They appear
if one multiplies the full tree level width, which contains for instance 
subleading contributions from the phase space integration and 
from the longitudinal
vector bosons, by the radiative correction 
factor
calculated in the leading $m_H$ approximation.
These terms are of 
the same order in the coupling constant as the theoretical uncertainty 
related 
to the use of the equivalence theorem while calculating radiative 
corrections.
It is thus not possible to decide unambiguously whether it is better 
to keep
them or to drop them without calculating the complete subleading 
contributions
explicitly. Numerically, this ambiguity is small and can be safely neglected.

The structure of the heavy Higgs radiative corrections to
the Higgs decay width into fermions and into vector bosons
is shown in fig. 3. Namely, the ratio of the decay widths including the
${\cal O}(\lambda)$ and ${\cal O}(\lambda^2)$ radiative corrections to the
tree level widths is plotted as a function of the Higgs mass. 
It should be remembered that the $m_H$ parameter is the
on--shell Higgs mass as defined by the renormalization conditions
of eqns. 2.

\begin{figure}
\hspace{.5cm}
    \epsfxsize = 14cm
    \epsffile{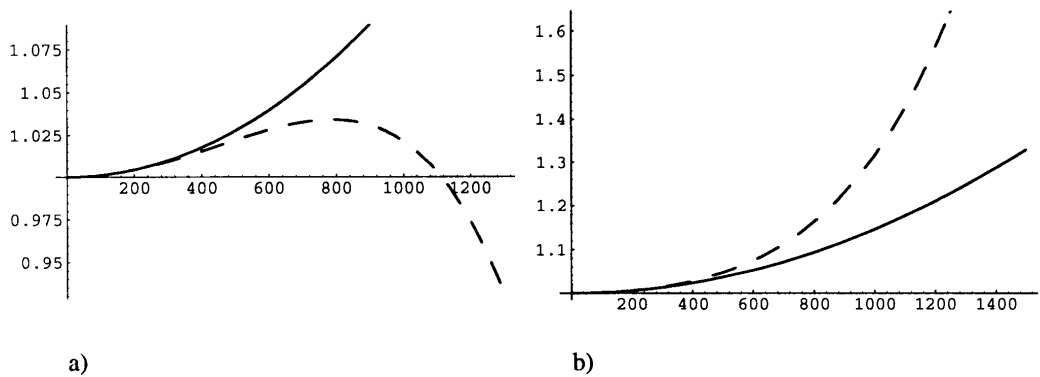}
\caption{The magnitude of the leading $m_H$ radiative corrections to
         the $H \rightarrow t\bar{t}$ (a) and 
         the $H \rightarrow ww$ (b) decays. The plots show the ratios
         of the decay widths at one--loop (solid line) and two--loop
         (dashed line) to the tree level decay widths as a function of
         the on--shell Higgs mass.}
\end{figure}

In the case of the $H \rightarrow t \bar{t}$ decay, the one--loop and the
two--loop corrections have opposite signs and therefore partly 
compensate each other. For a Higgs mass $m_H \sim 1.1$ TeV the two--loop 
correction becomes as large as the one--loop contribution. This is an indication
of the validity range of perturbation theory in the on--shell renormalization
scheme. The perturbative series is at best asymptotic. Its use is motivated
by the assumption that its first few terms display a reasonable 
convergence towards the unknown exact solution. If already the 
two--loop correction is as large as the one--loop one, the series 
appears to show no sign of convergence at all, 
and the validity of the perturbative approach becomes questionable.
This criterion for the breakdown of the perturbation theory 
was used previously by van der Bij and Veltman \cite{vdbij:rho} in the case
of the heavy Higgs contributions to the $\rho$ parameter, and by van der Bij
for the heavy Higgs corrections to the trilineal vector boson couplings 
\cite{vdbij:vertex}. They derived bounds
on the Higgs mass as heavy as 3---4 TeV because of the screening of heavy Higgs
effects. For the fermionic Higgs decay no screening is present, and
in this case the corresponding bound is considerably lower. One also notices that the sum 
of the one-- and two--loop radiative corrections is quite small 
over the whole range of validity of perturbation theory up to about 1.1 TeV.
Considering also the smallness of the $t\bar{t}$ branching ratio of heavy 
Higgses, this makes these effects quite marginal from a phenomenological point
of view.

The situation is different with the Higgs decay into vector bosons.
The two--loop correction becomes larger than the one--loop contribution 
for a Higgs boson mass larger than $\sim 930$ GeV. The one-- and two--loop 
corrections have the same sign and result in an enhancement of the decay
width with respect to the tree level. At $m_H \sim 930$ GeV the one--loop 
correction is still rather small, at 13\% level. The one--loop correction 
becomes numerically large only for considerably heavier Higgses, of the 
order of 1.3 TeV, as it was noticed in ref. \cite{marciano}. 
Still, the perturbation theory breaks down 
for a Higgs mass larger than about 930 GeV in the OMS scheme.
This is an interesting result which shows that a perturbative solution may be 
unreliable even if the one--loop radiative corrections are numerically small.

Before concluding this section, I would like to comment briefly on some
speculations about the possible relevance of the perturbative result
for a large Higgs mass, where the perturbative series diverges very badly.
In an attempt to extend the perturbative result in this zone, one can try to 
construct a diagonal sequence of Pad\'e approximants, as pointed out in
ref. \cite{2loop:Htott}. The hope is that this would sum up the asymptotic series, but
of course there is no formal proof that this procedure converges.
However, these speculations are encouraged by a relation 
which exists at least at leading order between
the Pad\'e approximants and the nonperturbative 1/N expansion of the $O(N)$
sigma model \cite{willenbrock}. In the case of the fermionic Higgs decay, 
the [1/1] Pad\'e approximant is a well behaved function which tends 
to a constant as the Higgs mass is increased. Still, in the case of the 
Higgs decay into vector boson pairs the [1/1] Pad\'e approximant has a pole
for a finite value of the Higgs mass, and so the relevance of
the Pad\'e approximant approach is not clear.


\section{Heavy Higgs searches at hadron colliders}

For $m_H \sim 930$ GeV, above which perturbation theory breaks down totally
at least in the OMS renormalization scheme, the total one-- and two--loop 
radiative corrections 
to the Higgs decay into vector bosons are quite substantial, 
of the order of 26\%. This leaded us to consider 
the r\^{o}le of this type of effects in other processes of interest in view 
of heavy Higgs searches at future colliders. In particular, 
it would be interesting to investigate the relevance of radiative
corrections of enhanced electroweak strength in Higgs production 
by gluon fusion and subsequent decay into vector bosons. 

The Higgs boson can be produced by gluon fusion via a heavy quark
loop. This  
processes is of special interest for Higgs searches at the LHC.
It was studied extensively at leading order, and the next--to--leading 
order QCD corrections were calculated by M. Spira et al. \cite{spira}.
Here we are interested in the next--to--next--to--leading order
radiative corrections of enhanced electroweak strength to this process.

\begin{figure}
\hspace{2.5cm}
    \epsfxsize = 10cm
    \epsffile{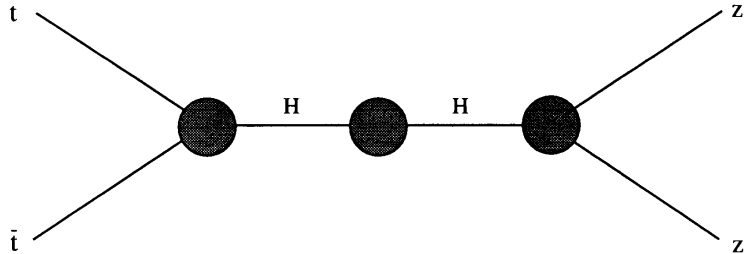}
\caption{The structure of the radiative corrections of enhanced electroweak 
         strength to the $t \bar{t} \rightarrow H \rightarrow zz$ 
         scattering process. No box diagrams contribute in the order
         considered here because they are of higher order in the quark mass.}
\end{figure}

Before calculating the NNLO corrections to the 
$gg \rightarrow H \rightarrow zz$ process, it is useful to consider 
first the related $t\bar{t} \rightarrow H \rightarrow zz$ scattering.
The structure of the leading $m_H$ radiative corrections to this process
is shown in fig. 4. The only contributions which need to be considered are
the corrections to the Higgs propagator and the corrections to the $Ht\bar{t}$ 
and $Hzz$ vertices. No other diagrams can contribute at the order considered here. 
For instance, box diagrams are of higher order in the top quark mass.

In order to calculate the radiative corrections to this scattering process
consistently as an expansion in the coupling constant, one needs to pay
special attention to the treatment of the Higgs resonance. The Dyson summation
introduces inverse powers of the coupling constant. As a result, for 
deriving the complete NNLO corrections to the 
$t\bar{t} \rightarrow H \rightarrow zz$ process in the resonance region, 
one needs to include the two--loop corrections to the Yukawa coupling and to the
$Hzz$ coupling, and the Higgs selfenergy up to three--loop. In fact, only the
imaginary part of the three--loop Higgs selfenergy is needed at the order
considered here, and this can be calculated from the two--loop Higgs decay
into a pair of vector bosons and from the tree level Higgs decay into four
vector bosons. The details of the calculation can be found in 
ref. \cite{glufusion}.
By taking into account all relevant contributions, one obtains the full
NNLO corrections to the shape of the Higgs resonance which are shown in fig. 5.

\begin{figure}
\hspace{2.5cm}
    \epsfxsize = 10cm
    \epsffile{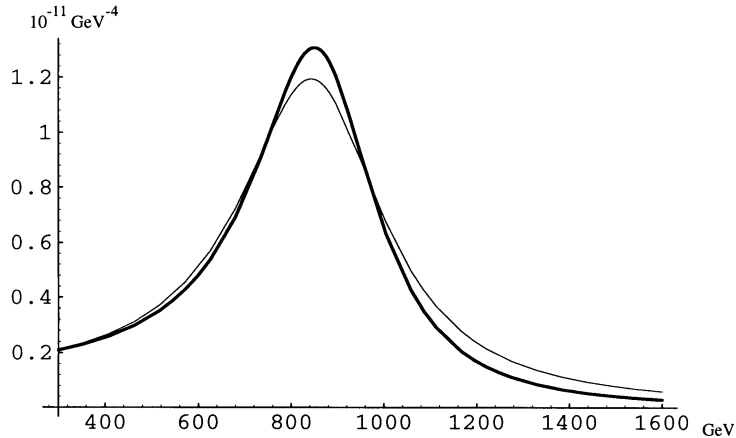}
\caption{The effect of the NNLO corrections of enhanced electroweak strength 
         on the shape of the Higgs resonance in the
         $t\bar{t} \rightarrow H \rightarrow zz$ scattering for 
         $m_H = 850$ GeV. The solid line is the tree level 
         and the thin line is the NNLO result.}
\end{figure}

At this point one can calculate the corrections to the gluon fusion process.
Apart from the triangular Higgs production diagram, there are also
background box diagrams which contribute to the $gg \rightarrow ZZ$ process,
as shown in fig. 6. The two types of diagrams behave differently as a function
of the quark mass. The triangle diagram results in an effective $Hgg$ coupling 
in the heavy quark limit, while the box diagrams decouple. The leading
$m_H$ correction to the triangle diagram are the same as those derived for the 
$t\bar{t} \rightarrow H \rightarrow zz$ scattering, and are independent 
of the top mass. The box diagrams can receive corrections from
the rescattering of the outgoing vector bosons. These corrections are formally
of order $\lambda$, but they depend on the precise ratio of the top 
and Higgs masses. Because the $t\bar{t}$ threshold of $\sim 360$ GeV is 
not negligible with respect to the Higgs mass, which will be taken of the 
order of 700---900 GeV, an expansion in the the top mass will probably be  
a not very useful approximation, and the full dependence on the top mass would 
need to be taken into account in these diagrams. 
This type of combined top--Higgs mass corrections 
may be numerically relevant for heavy Higgs searches at hadron colliders.
Technically, their evaluation is difficult because one needs to calculate 
two--loop box diagrams and even three--loop vertex diagrams. 
Here I will consider
only the universal corrections to the triangle diagram, which are independent
of the top mass. This approximation is exact in the heavy top limit, when the 
box diagrams decouple. 

\begin{figure}
\hspace{1.5cm}
    \epsfxsize = 12cm
    \epsffile{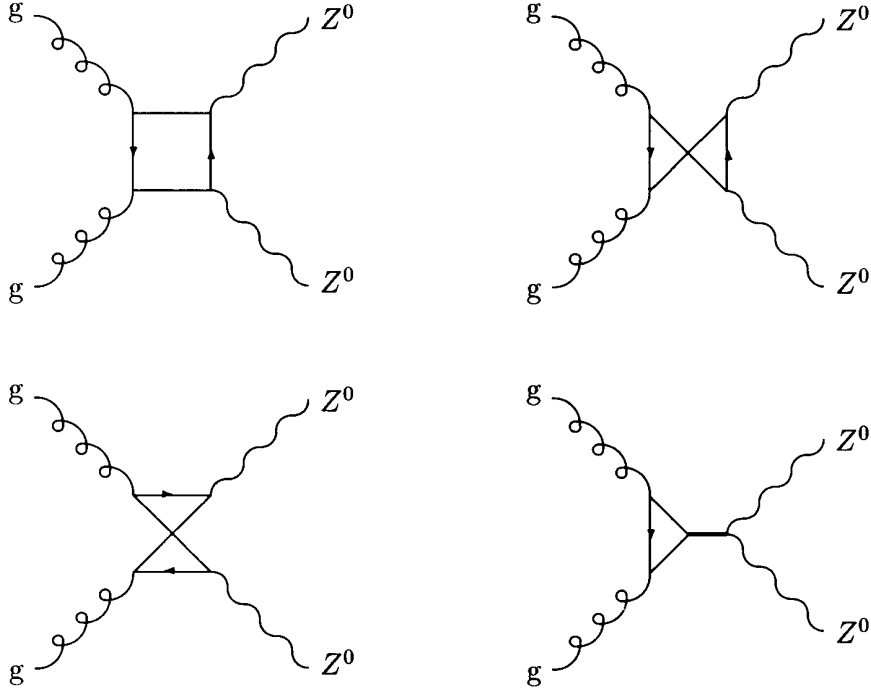}
\caption{The leading order diagrams which contribute to the 
         $gg \rightarrow ZZ$ process.}
\end{figure}

We have incorporated the NNLO radiative corrections of 
enhanced electroweak strength in a Monte Carlo simulation of the 
Z pair production at the LHC. The details of the calculation can be found
in ref. \cite{glufusion}. The results of the simulation are shown in figs. 7 and 8. 
Comparing with the $t\bar{t} \rightarrow H \rightarrow zz$ process, one notices 
that the effect of the radiative corrections in gluon fusion is an enhancement
of the cross section because of interference effects with the box diagrams.
This enhancement is at the level of 10---20\%, depending on the mass of the 
Higgs boson.

\begin{figure}
\hspace{1.5cm}
    \epsfxsize = 12cm
    \epsffile{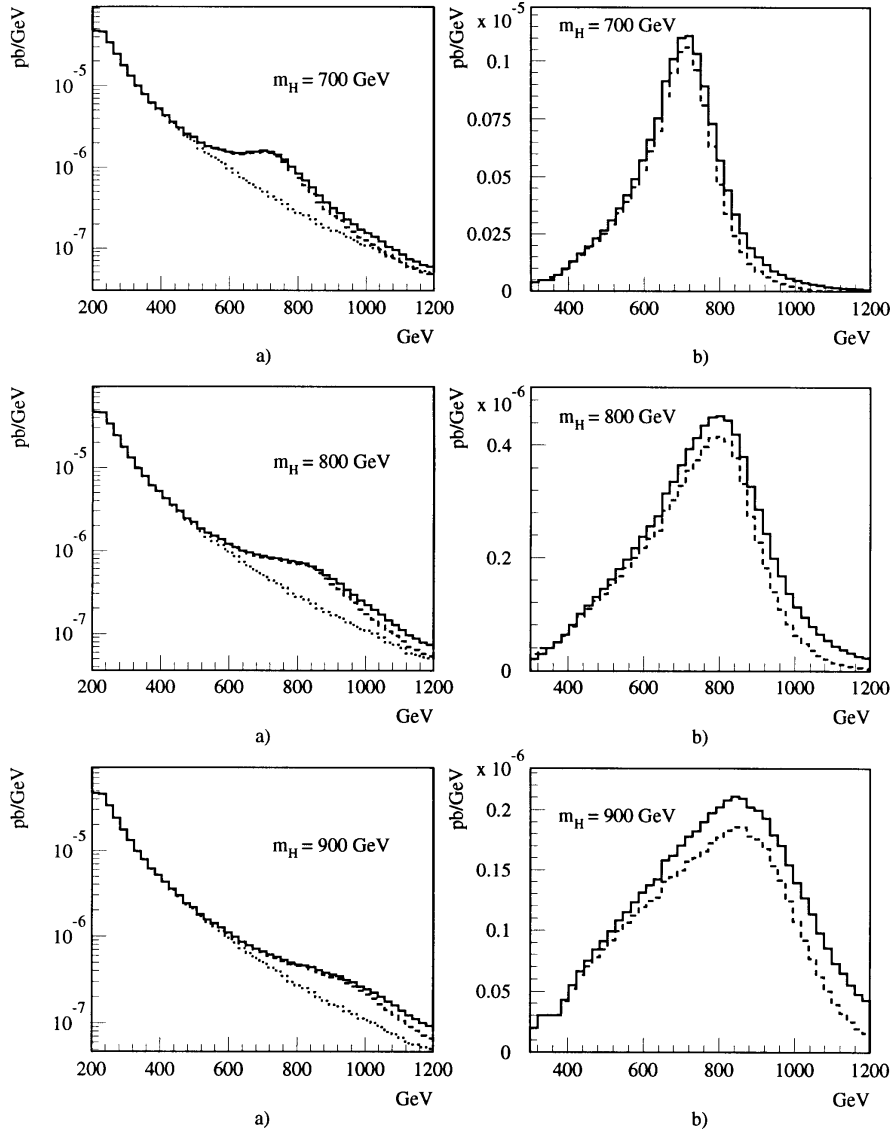}
\caption{Invariant mass distribution of the $Z^0$ pairs at LHC.
               The processes considered are 
	       $gg \rightarrow ZZ \rightarrow 2(\mu^+\mu^-)$
	       and
	       $q\bar{q} \rightarrow ZZ \rightarrow 2(\mu^+\mu^-)$.
	       We consider a CM energy of 14.5 TeV, and for the outgoing
	       muons we request $p_T>20$ GeV and $|y_l|<2.5$.
               The solid line is the NNLO cross section, the dashed
	       line is the tree level cross section, and the dotted line
	       is the background (no Higgs production diagram).
	       a) shows the total cross section, and b) shows the
	       Higgs signal, with the background subtracted.}
\end{figure}

\begin{figure}
\hspace{1.5cm}
    \epsfxsize = 12cm
    \epsffile{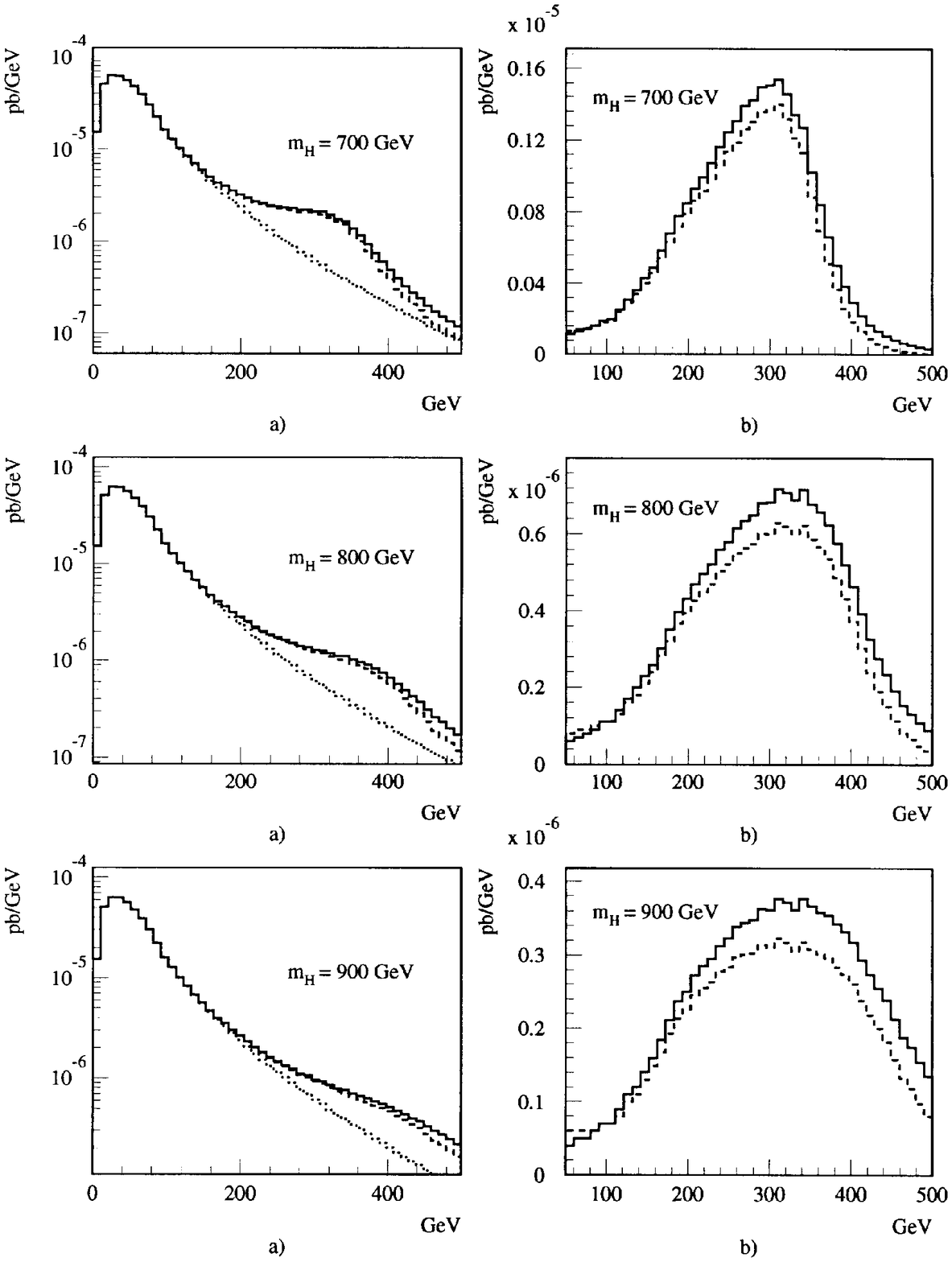}
\caption{Transverse momentum distribution of the $Z^0$ bosons.
         Same as fig. 7.}
\end{figure}


\section{Conclusions}

Higher order radiative corrections of enhanced electroweak strength
become increasingly important as the mass of the Higgs boson is increased.
They are interesting phenomenologically in view of Higgs searches at future
colliders. They also provide insight in the breakdown of perturbation
theory as the Higgs selfinteraction becomes strong. 

The calculations in the Higgs sector beyond one--loop level are challenging 
because they involve the evaluation of massive diagrams. A powerful technique 
is available which allows one to deal with any two--loop diagram. Similar
methods were developed for a class of three--loop diagrams as well.

These techniques were used for calculating a number of processes involving 
the Higgs sector of the standard model at two--loop level. This allows one
to set perturbative bounds on the mass of the Higgs particle, beyond which
the perturbative approach is not reliable anymore.

An interesting point is that perturbation theory may cease to be reliable already
for values of the coupling for which the one--loop corrections are still rather 
small, as it was shown explicitly in the case of the Higgs decay into vector 
bosons.

Finally, the analysis of the two--loop heavy Higgs effects in the Higgs boson 
production by gluon fusion shows that this type of effects may be numerically 
important for heavy Higgs searches at the LHC.

An interesting point which I did not discuss is the longitudinal 
vector boson scattering. This process shows promise of providing
insight in the spontaneous electroweak symmetry breaking mechanism,
and becomes important as a source of Higgs bosons at hadron colliders
for $m_H \sim 1$ TeV. This process was studied at one--loop order
in ref. \cite{dawson,yndurain}. At two--loop level only a calculation in the 
high energy limit exists, where the Feynman diagrams which are involved 
are simpler \cite{maher}. A complete two--loop analysis would be difficult 
because it involves the calculation of two--loop massive box diagrams. This
is an interesting problem which deserves further investigation.


\vspace{.5cm}

{\bf Acknowledgement}

I would like to thank the theory department of the Brookhaven National
Laboratory, where this paper was written, for hospitality, and the
U.S. Department of Energy (DOE) for support.
This work was supported by the Deutsche Forschungsgemeinschaft (DFG).



\end{document}